\documentclass[preprint]{elsarticle}
\usepackage{lineno,hyperref}
\modulolinenumbers[5]
\journal{}
\usepackage{setspace}

\usepackage{geometry}

\newgeometry{vmargin={25mm}, hmargin={20mm,20mm}}
\usepackage{color}
\usepackage{amsmath}

\usepackage{lineno}
\usepackage{comment}
\usepackage{setspace}
\usepackage{geometry}
\usepackage{color}
\usepackage{amsmath}
\usepackage{caption}
\usepackage{subcaption}
\begin{document}

\begin{frontmatter}

\title{Dynamical analysis of the infection status in diverse communities due to COVID-19 using a modified SIR model}

\author[IC_mainaddress]{Ian Cooper}

\author[AM_mainaddress,CA_mainaddress]{Argha Mondal}

\author[CA_mainaddress]{Chris G. Antonopoulos\corref{CAcorrespondingauthor}}
\cortext[CAcorrespondingauthor]{Corresponding author}
\ead{canton@essex.ac.uk}

\author[AMi_mainaddress]{Arindam Mishra}

\address[IC_mainaddress]{School of Physics, The University of Sydney, Sydney, Australia}
\address[AM_mainaddress]{Department of Mathematics, Sidho-Kanho-Birsha University, Purulia-723104, WB, India}
\address[CA_mainaddress]{Department of Mathematical Sciences, University of Essex, Wivenhoe Park, UK}
\address[AMi_mainaddress]{Division of Dynamics, Technical University of Lodz, Stefanowskiego 1/15, 90-924 Lodz, Poland}

\begin{abstract}
In this article, we model and study the spread of COVID-19 in Germany, Japan, India and highly impacted states in India, i.e., in Delhi, Maharashtra, West Bengal, Kerala and Karnataka. We consider recorded data published in Worldometers and COVID-19 India websites from April 2020 to July 2021, including periods of interest where these countries and states were hit severely by the pandemic. Our methodology is based on the classic susceptible-infected-removed (SIR) model and can track the evolution of infections in communities, i.e., in countries, states or groups of individuals, where we (a) allow for the susceptible and infected populations to be reset at times where surges, outbreaks or secondary waves appear in the recorded data sets, (b) consider the parameters in the SIR model that represent the effective transmission and recovery rates to be functions of time and (c) estimate the number of deaths by combining the model solutions with the recorded data sets to approximate them between consecutive surges, outbreaks or secondary waves, providing a more accurate estimate. We report on the status of the current infections in these countries and states, and the infections and deaths in India and Japan. Our model can adapt to the recorded data and can be used to explain them and importantly, to forecast the number of infected, recovered, removed and dead individuals, as well as it can estimate the effective infection and recovery rates as functions of time, assuming an outbreak occurs at a given time. The latter information can be used to forecast the future basic reproduction number and together with the forecast on the number of infected and dead individuals, our approach can further be used to suggest the implementation of intervention strategies and mitigation policies to keep at bay the number of infected and dead individuals. This, in conjunction with the implementation of vaccination programs worldwide, can help reduce significantly the impact of the spread around the world and improve the wellbeing of people.
\end{abstract}


\end{frontmatter}

%\linenumbers

\section{Introduction}\label{sec_intro}

A novel strand of Coronavirus (SARS-CoV-2) was identified in Wuhan, Hubei Province in China in December 2019 that causes a severe and potentially fatal respiratory syndrome, i.e., COVID-19. Since then, it has become a pandemic declared by the World Health Organization (WHO) on 11 March 2020 and has spread around the globe \cite{tang2020estimation,wu2020new,Who2020}. The coronavirus disease \cite{wu2020new} due to COVID-19, has impacted heavily on the human health and the socioeconomic status in affected countries. Governments and local authorities have no choice other than taking diverse and adequate countermeasures due to still limited information about COVID-19. Until now, various countermeasures have been proposed and implemented, such as wearing face masks, sanitization, following social-distancing, implementing lockdowns and quarantines, etc. Recently, vaccination programs have started springing-up around the world \cite{Who2021}, and even though there is early success, there are still challenges remaining \cite{Bar-Zeev2020}.

Recent, modelling and experimental studies \cite{wu2020new,wu2020nowcasting,ranjan2020predictions,Who2020,iwasaki2021reinfections,boccaletti2020modeling,machado2020nonlinear} on the changes of susceptibility, infection rates, deaths and recovered cases from COVID-19 can help governments and local authorities implement countermeasures to reduce the infection rates substantially. For example, in India, various reactive measures have been taken in various states after the number of infected cases soared \cite{bagal2020estimating,MALAVIKA202126,basu2020going,das2021covid,pai2020investigating,rafiq2020evaluation,samui2020mathematical,sarkar2020modeling}. It is known that early in the pandemic and after the onset of secondary waves, if measures are not taken to mitigate the spread, the number of infected cases grows exponentially fast with a certain transmission rate. However, this can change due to public awareness and countermeasures implemented by governments and local authorities. Hence it becomes imperative the necessity to explain and forecast the future trajectory of the spread of COVID-19 to help governments and authorities make decisions to implement timely countermeasures, control strategies and allocate wisely financial and medical resources.

There are many recent research articles available online that predict the development trend of the pandemic in various countries and regions \cite{liao2020tw,liu2021predicting}. In our analysis, we use recorded data sets published in \cite{Worldometer_website,COVID19-India_website} for India, Japan and Germany and the Indian states of Delhi, Maharashtra, West Bengal, Kerala and Karnataka, from April 2020 to July 2021, including periods of interest where these countries and states were hit severely by the pandemic. Our approach is based on a modification of the classic SIR model \cite{weiss2013SIR} and is motivated by our earlier research on the spread of COVID-19 in different communities in \cite{cooper2020sir,cooper2020dynamic} and \cite{liao2020tw}. The number of infections over time in Japan exhibits a wavy pattern, pointing to the onset of secondary waves or surges and various types of infection curves have been observed during the spread of the virus \cite{kobayashi2020predicting,dashtbali2021compartmental,odagaki2020analysis,odagaki2021classification}. Monitoring these onsets, surges, outbreaks or secondary waves and depending on the infection and transmission rates, governments and local authorities can decide to impose a range of measures to mitigate or slow down the spread of the virus.

In the context of the current situation worldwide, appropriate epidemic models \cite{liao2020tw,liu2021predicting,giordano2020modelling,postnikov2020estimation,fanelli2020analysis,scarpino2019predictability,mishra2021mathematical,munoz2021sir,tyagi2021mathematical} for the prediction of the spread of COVID-19 in different countries and communities are highly relevant and important. In particular, the SIR model and its extended modifications \cite{Hethcote1989,Hethcote2000,Hethcote2008,Weiss2013}, such as the extended SIR model in various forms have been used in previous studies \cite{LOPEZ2021103746,Ndairou2020} to model the spread of COVID-19 in communities. Hence forecasting using infectious disease models is a widely used approach, including the ongoing pandemic. However, these models depend on various assumptions and different conditions. Among those models, probably the most used one is the classic SIR model that dates back to the work by R. Ross, W. Hamer, and others in the early twentieth century \cite{Anderson1991,weiss2013SIR}, which consists of a system of three coupled ordinary differential equations. In our work, oscillations in the solutions to the modified SIR model have also been described by suitable choices of model parameters estimated from the recorded data sets.

The transmission of COVID-19 from infected to susceptible individuals depends upon diverse influences and factors, and thus can be challenging to understand. However, using our methodology, we show that by tracking the spread of the virus on a regular (e.g., daily) basis, possible future scenarios can be explored. The main characteristics that describe the spread of COVID-19 can easily be understood, without making use of complex assumptions. Comparing the model predictions of our approach with the recorded data sets, one can assess the effectiveness of implemented measures to control or mitigate the spread of the virus and forecast the trajectory of the spread in communities.

The model we propose here is based upon individuals within a community grouped into three compartments or populations: the susceptible population whose individuals can become infected, the infected population and the removed population  who have either recovered or died due to the virus. We begin by estimating the infection rates using the proposed model and then computing the infection and other rates from the recorded data sets. Our approach is designed to be flexible as it can be updated in real-time as new data are coming in. As they become available, they are added to a database and are used to update the input parameters to get manually a best-fit between them and the model predictions for the susceptible, infected and recovered populations. It is this inherent flexibility that renders our approach an excellent forecaster for the spread of the virus in communities, either being countries, states or groups of individuals. Our work has the potential to provide information that can help understand how the virus spreads in communities and can offer governments and authorities insights into it, when and how to introduce policies to keep the spread of the virus under control.

It is worth it to mention that the proposed model is adjusted by estimating the model parameters manually to best match the population plots between model predictions and recorded datasets. It would be possible to automate the fit by some least squares approach, however it is not necessary since one is fitting predictions to possibly poor quality data. Importantly, the model we propose here does not describe a single wave. It can track the past reasonably well and one can make predictions about future waves which can be used to test a variety of scenarios.

The paper is organized as follows: In Sec. \ref{sec_methodology}, we introduce our mathematical modeling approach based on the modification of the classic SIR model, discuss its various aspects and explain the approach in Sec. \ref{sec_results} to study the recorded data sets in \cite{Worldometer_website} for Germany, Japan, India and in \cite{COVID19-India_website} for some select key-states in India. In Subsec. \ref{subsec_forecasting}, we study the future trajectory of the spread of the virus in Japan and India and assess the results that stem from our analysis. Finally, in Sec. \ref{sec_discussion}, we conclude our work and discuss possible outcomes of our analysis and its connection to the evidence that has been already collected on the spread of COVID-19 worldwide.

%%%%%%%%%%%%%%%%%%%%%%%

\section{Methodology}\label{sec_methodology}

\begin{figure}[!ht]
\centering
\begin{subfigure}[b]{0.48\textwidth}
\centering
\includegraphics[scale=0.25]{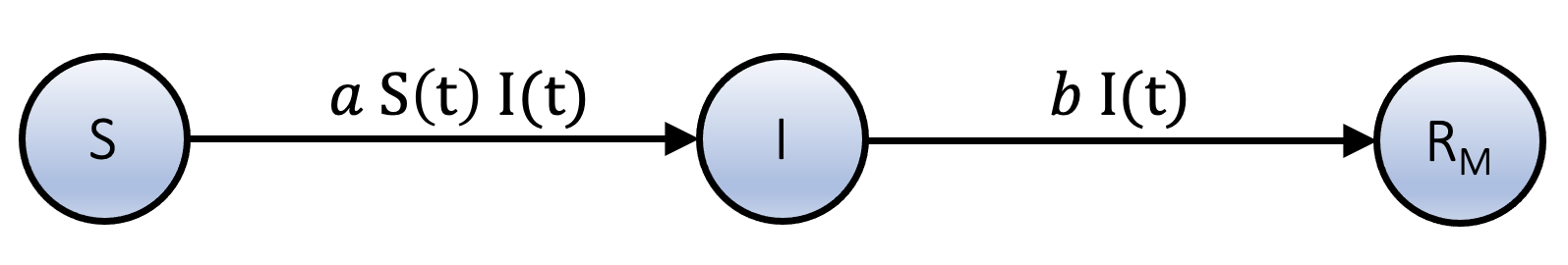}
\caption{}
\end{subfigure}
\hfill
\begin{subfigure}[b]{0.48\textwidth}
\centering
\includegraphics[scale=0.25]{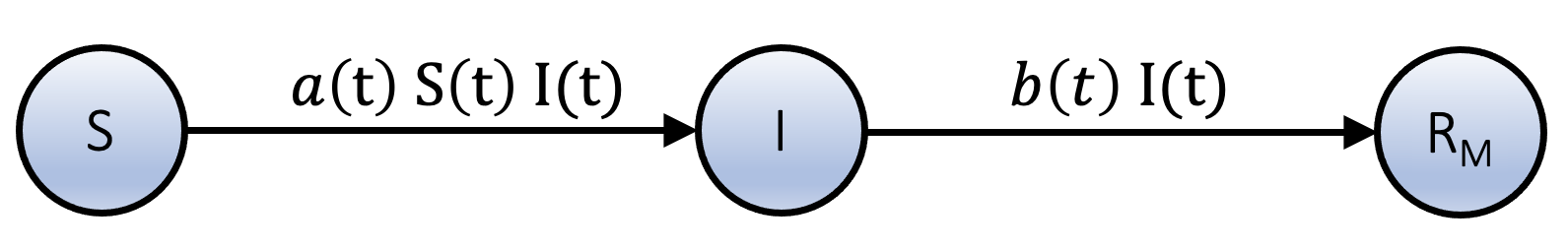}
\caption{}
\end{subfigure}
\hfill
\begin{subfigure}[b]{0.54\textwidth}
\centering
\includegraphics[scale=1]{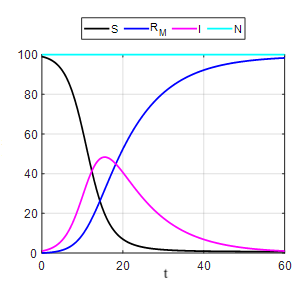}
\caption{}
\end{subfigure}
\caption{The population-flux diagrams of the SIR models \eqref{eq_classic_SIR_model} and \eqref{eq_modified_SIR_model}, and an example of numerical solutions to the classic SIR model \eqref{eq_classic_SIR_model}. Panels (a) and (b) show how the susceptible, $S$, infected, $I$ and removed, $R_M$, populations interact. Parameter $a$ is the effective transmission rate and indicates that each susceptible individual infects randomly $a$ individuals every day and $b$ the recovery rate and indicates that the infected individuals recover or die with probability $b$. In panel (a), $a$ and $b$ are constants. In panel (b) and model \eqref{eq_modified_SIR_model}, $a$ and $b$ are functions of time $t$ and are estimated manually so that the rescaled $I$ from the solution to system \eqref{eq_modified_SIR_model} and rescaled $I^d$ from the recorded data set match closely. The arrows show the direction of the flow of the populations $S$, $I$ and $R_M$. Panel (c): An example of numerical solutions to the classic SIR model \eqref{eq_classic_SIR_model} using the fourth-order Runge-Kutta method: Time evolution of the susceptible, $S$, (in black), removed, $R_M$, (in blue), infected, $I$, (in light blue) and total, constant, population, $N$, (in magenta). The susceptible population, $S$, in black decreases quickly to zero as more and more individuals become infected, $I$ (in light blue). At the same time, the removed population, $R_M$, increases steadily, converging to the total population, $N$. Here the total population, $N$, is 100 and initially (at $t$=0), the whole population is considered susceptible to the virus, hence $S(0)=100$ and $I(0)=R_M(0)=0$.}\label{Fig1}
\end{figure}

The classic SIR model is a simple compartmental model routinely used as an epidemic model \cite{weiss2013SIR,Kermacketal2020,amaro2021global}. It consists of the three coupled ordinary differential equations given in system \eqref{eq_classic_SIR_model} that describe the evolution of the susceptible $S$, infected, $I$, and removed, $R_M$, populations over time $t$ from a total, constant, population $N=S+I+R_M$. In particular, the total, constant, population $N$ is divided into the following three populations (or compartments): 
\begin{enumerate}
\item {Susceptible population, $S(t)$: These are those individuals who are not infected, however, could become infected. A susceptible individual may become infected or remain susceptible. As the virus spreads from its source or new sources spring up, more individuals  become infected, thus the susceptible population will decrease in time.}
\item{Infected population, $I(t)$: These are those individuals who have already been infected by the virus and can transmit it to the susceptible individuals. An infected individual may remain infected, and can be removed from the infected population to recover or die.}
\item{Removed population, $R_M(t)$: These are those individuals who have either recovered from the virus and are assumed to be immune, $R_C(t)$ or have died, $D(t)$, thus $R_M=R_C+D$.}
\end{enumerate}
Furthermore, it is assumed that the time scale of the SIR model is short enough so that births and deaths, other than deaths caused by the virus, can be neglected and that the number of deaths from the virus is small compared with the living population. The SIR model is given by the system of ordinary differential equations
\begin{align}\label{eq_classic_SIR_model}
\frac{dS(t)}{dt}&=-aS(t)I(t),\nonumber\\
\frac{dI(t)}{dt}&=aS(t)I(t)-bI(t),\\
\frac{d{R_M}(t)}{dt}&=bI(t),\nonumber
\end{align}
where $a$ is the effective transmission rate and indicates that each susceptible individual infects randomly $a$ individuals every day and $b$ is the recovery rate and indicates that the infected individuals recover or die with probability $b$. In the context of the classic SIR model \eqref{eq_classic_SIR_model}, $a$ and $b$ are constants and the recovered, $R_C$, or dead, $D$, individuals cannot be distinguished, so they are represented by $R_M$. The population-flux diagram of the SIR model \eqref{eq_classic_SIR_model} which shows how $S$, $I$ and $R_M$ interact can be seen in Fig. \ref{Fig1}(a). The model is derived using several assumptions. First, it is assumed that the members of the susceptible and infected populations are homogeneously distributed in space and time. Then, an individual removed from the infected population has lifetime immunity and that the total population $N$ is constant in time. Finally, it is also assumed that the number of births and deaths from causes other than the virus are ignored.

Since the equations of the classic SIR model \eqref{eq_classic_SIR_model} comprise a system of coupled ordinary differential equations, finding analytical solutions in closed form using known mathematical functions is difficult. The process to find them is complicated and there are limitations in practical applications \cite{Harkoetal2014}. Hence a common approach is to solve system \eqref{eq_classic_SIR_model} numerically. Here, we opted for the classic fourth-order Runge-Kutta numerical integrator to obtain the approximated solutions to $S$, $I$ and $R_M$ at discrete time steps $dt=0.1$. In our model, $t$ is expressed in days. A set of numerical solutions to system \eqref{eq_classic_SIR_model} is shown in Fig. \ref{Fig1}, where initially, the time evolution of the infected population, $I$, (in light blue) is observed to increase until it reaches its peak value, after which it decreases monotonically to zero. The susceptible population, $S$, (in black) decreases quickly to zero as more and more individuals become infected or removed. At the same time, the removed population, $R_M$, (in blue) increases steadily converging to the total population, indicating that all individuals have been removed because they have either recovered from the virus and are assumed immune or have died. In this example, the total population $N$ is 100 and initially (i.e., at $t$= 0), the whole population is considered susceptible to the virus, hence $S(0)=100$ and $I(0)=R_M(0)=0$.

Undoubtedly, model \eqref{eq_classic_SIR_model} is simple and useful to study the spread of viruses in closed communities when there is only one outbreak of infections. However, it cannot describe the spread of COVID-19 completely  when there are more than one outbreaks or surges or when the system is not closed, i.e., when the total population $N$ is not constant in time. This is indeed the case with the recorded data sets in \cite{Worldometer_website} for India, Japan and Germany and in \cite{COVID19-India_website} for the Indian states that we study here. Thus, to model the spread of COVID-19 in such cases, we consider the  modified SIR model
\begin{align}\label{eq_modified_SIR_model}
\frac{dS(t)}{dt} &=-a(t)S(t)I(t),\nonumber\\
\frac{dI(t)}{dt} &= a(t)S(t)I(t)-b(t)I(t),\\
\frac{d{R_M}(t)}{dt} &= b(t)I(t),\nonumber
\end{align}
%where $R_M = R_C + D$ is the population of individuals removed from the infected population, $Ix$ and can have either recovered, $R_C$ or died $D$. In the following, we will discuss a way to estimate $R_C$ and $D$ from the recorded data sets and the solutions to system \eqref{eq_modified_SIR_model}. Consequently, the numerical solutions for the active infections $I$ and removals $R_M$ are multiplied by two population, scaling-factors $f_1$ and $f_2$ that are estimated manually so that the rescaled $I$ and $R_M$ solutions match closely to the variables $I^d$ and $R^d$ of the recorded data sets, that we discuss next.
which is solved using scaled values for the initial conditions $I(0)$ and $R_M(0)$ and an initial condition $S(0)$ in $[0,1]$. In particular, $I=fI^{us}$ and $R_M=fR_M^{us}$, where $I^{us}$, $R_M^{us}\in[0,1]$. The scaling-factor $f$  and $I^{us}(0)$, $R_M^{us}(0)\in[0,1]$ are estimated manually so that $I(t)$ and $R_M(t)$ match in time closely the variables $I^d$ and $R^d$ of the recorded data sets, that we discuss next. This makes it easier to reset the values of $S(t)$ and $I(t)$ at a time $t$ as described by Eq. \eqref{eq_reset_S}. In this framework, the removed population, $R_M$, consists of those individuals who have recovered, $R_C$, or have died, $D$, hence $R_M(t) = R_C(t) + D(t)$. The population-flux diagram of model \eqref{eq_modified_SIR_model} is shown in Fig. \ref{Fig1}(b), where $a$ and $b$ are no longer constants, but functions of time $t$.

The recorded data sets in \cite{Worldometer_website} and \cite{COVID19-India_website} are organized in the form of time-series where the rows are recordings in time from April 2020 to July 2021, and the columns are the total infections due to the virus, $I^d_{tot}$, number of infected individuals, $I^d$, and deaths, $D^d$. Importantly, they may exhibit spikes in the number of infected individuals, $I^d$, at specific times $t_i$ (i.e., more than one outbreaks of secondary waves or surges), which result in the increase of the susceptible population, $S$ when they occur, thus the total population cannot be considered constant in the classic SIR model in system \eqref{eq_classic_SIR_model}. To account for this, every time there is an outbreak in the number of infected individuals, $I^d$ with the recorded data set in \cite{Worldometer_website} or \cite{COVID19-India_website}, we reset $S$ and $I$ in system \eqref{eq_modified_SIR_model} by
\begin{align}
S(t_i)&=S(t_i)+\Delta S(t_i)\label{eq_reset_S},\\
I(t_i)&=I(t_i)+\Delta I(t_i)\label{eq_reset_I},
\end{align}
where $t_i$ are the times where the outbreaks occur in the recorded data sets and $i$ the index that runs through the $M$ outbreaks in the recorded data set. Here, the susceptibility-factor increment, $\Delta S$, and infection increment, $\Delta I$, are the reset values such that the (rescaled) solutions to $I$ and $R_M$ from system \eqref{eq_modified_SIR_model}, match closely $I^d$ and $R^{d}=I^d_{tot}-I^d$ from the recorded data set, respectively. Resetting $S$ and $I$ accounts for the mobility of individuals, i.e., when moving around communities. For example, travelers entering a country from overseas are often quarantined in hotels and this adds to the number of active infections in that community. As it is difficult to isolate the virus carried by infected individuals within, for example, hotel environments, the virus can escape from confined environments, increasing the number of infected and susceptible individuals in the wider community.

For the same reasons, we also assume that the effective infection rate, $a$ and recovery rate, $b$ in system \eqref{eq_modified_SIR_model} are functions of time $t$, rather than constants as in the classic SIR model \eqref{eq_classic_SIR_model}. We estimate $a(t)$ and $b(t)$ manually so that the rescaled $I$ from the solution to system \eqref{eq_modified_SIR_model} and rescaled $I^d$ from the recorded data set match closely. This approach allows to accommodate the different exponential growths and decays in the number of active infections, $I$, when there are multiple secondary outbreaks or surges in the recorded data sets. Hence our approach here is different than the classic SIR model \eqref{eq_classic_SIR_model}, which can only describe the initial exponential growth and decay in the spread of a virus.

Two of the most important variables in the spread of a virus in a community is the number of deaths, $D$, and recoveries, $R_C$. As these are not provided directly by model \eqref{eq_modified_SIR_model}, we can first estimate $D$ from the removals, $R_M$ of model \eqref{eq_modified_SIR_model} and the recorded data set and then, $R_C$, as follows: we define $M$, consecutive time-windows of interest $c_i=[t_i,t_{i+1})$, constants $\eta(c_i)$ and $k(c_i)$, where $i=1,\ldots,M-1$. Then, we estimate the deaths, $D$ in time by
\begin{equation}\label{eq_D}
D(c_i) = D(t_i) +\eta(c_{i+1})\big( {1-e^{- k(c_i) (R_M(c_i)-R_M(t_i))}} \big),
\end{equation}
where $M$ is the number of outbreaks or surges in the recorded data set. The time-windows of interest $c_i$ correspond to consecutive time-windows that contain the outbreaks or surges in infections. The idea behind Eq. \eqref{eq_D} is that typically, at the outbreaks of secondary waves or surges of infections, the rate at which people die is quite high at first, then steadily decreases to zero due to mitigation measures before another outbreak or surge occurs. We can then estimate the number of recovered individuals due to the virus, $R_C$ using the estimated $D(t)$ by
\begin{equation*}
R_C=R_M-D.
\end{equation*}
Finally, the number of total infections due to the virus, $I_{tot}$, can be estimated by
\begin{equation*}
I_{tot} = I + R_C + D.
\end{equation*}
Consequently, the total population $N$ is not defined for the modified SIR model \eqref{eq_modified_SIR_model} and hence $S$ does not represent the susceptible population but rather a susceptibility factor.

In the next section, we present the results of the application of the proposed methodology to data from Germany, Japan, India, Delhi, Maharashtra, West Bengal, Kerala and Karnataka as well as a forecast of the spread of COVID-19 in India and Japan.

%%%%%%%%%%%%%%%%%%%%%%%

\section{Results}\label{sec_results}

\subsection{Germany}\label{subsec_results_Germany}

\begin{figure}[!ht]
\centering
\includegraphics[width=\textwidth,height=11cm]{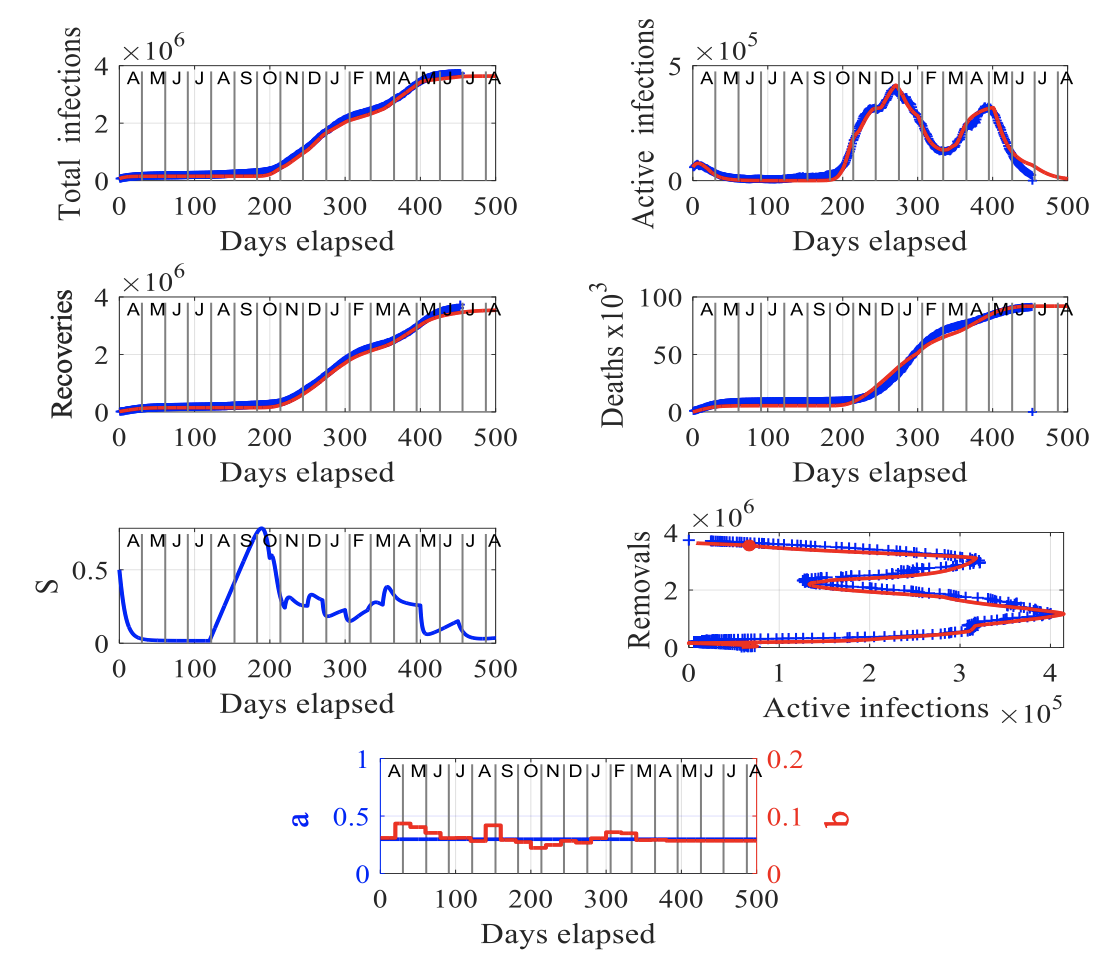}
\caption{Results for Germany: The input parameters to model \eqref{eq_modified_SIR_model} are $f = 1.7 \times 10^5$, $S(0) = 0.5$, $I(0) = 0.34324$ and $R_M(0) =0.11548$. The letters in the upper axes stand for the names of the months in a year, starting with ``A'' for ``April'' and the red curves are the solutions from our methodology, superimposed to the recorded data shown in blue. To obtain a good fit between the model's predictions and the recorded data, we have assumed that the rate parameter $b$ evolves with time as shown in the bottom plot.}\label{Fig_Germany}
\end{figure}

We start by studying the data set for Germany published in \cite{Worldometer_website}, which contains recordings between April 2020 and July 2021. Figure \ref{Fig_Germany} shows the evolution of $I_{tot}$, $R_C$, $S$, $a$, $b$, $I$, $D$ in time (days elapsed) and $R_M$ vs $I$. We can see in the plot of active infections, $I$, over time that there were two major, secondary waves of infections that started around the beginning of October 2020 and beginning of March 2021 in Germany, with the second being as high in the number of infections as almost the first. The number of total infections, $I_{tot}$ and recoveries, $R_C$, seem to stabilize in May and June 2021, in accordance with the (non-scaled) susceptibility factor, $S$ which shows a trend to decrease, fluctuating, to small numbers by August 2021. These results are further corroborated by the plot of deaths, $D$ over time, which again stabilize in June and July 2021.

The plot of removals, $R_M$, versus active infections, $I$, in Fig. \ref{Fig_Germany} is useful as the horizontal peak is a marker of a major outbreak in the country. Indeed, we can see that there are two horizontal peaks, pointing to the two outbreaks in Germany that peaked in late December 2020 and late April 2021. We also show in the figure, the parameters $a$ (in blue) and $b$ (in red) over time that appear in system \eqref{eq_modified_SIR_model} used to match closely the (rescaled) active infections, $I$, from the solution to system \eqref{eq_modified_SIR_model}, with the rescaled $I^d$ from the recorded data set. This allows to accommodate the different exponential growths and decays in the number of active infections, $I$, in the two outbreaks. The parameter $a$ (in blue) is constant throughout the 500 days worth of data, whereas $b$ shows fluctuations in time with a trend to converge after about March 2021.

Importantly, our modeling approach can provide valuable insights into the future trajectory of the spread of the virus shown in red in the plots in Fig. \ref{Fig_Germany}, and in particular, between July and August 2021, in the case where no major outbreaks will occur within that period. 

Clearly, our model-based analysis and forecast shows that the second outbreak in Germany is currently fading out and unless, another outbreak or something in favor of increasing the infection rate does not occur, the situation will improve, owing probably to the ongoing vaccination program in the country.

%%%%%%%%%%%%%%%%%%%%%%%%%%%%%%

\subsection{Japan}\label{subsec_results_Japan}

\begin{figure}[!ht]
\centering
\includegraphics[width=\textwidth,height=11cm]{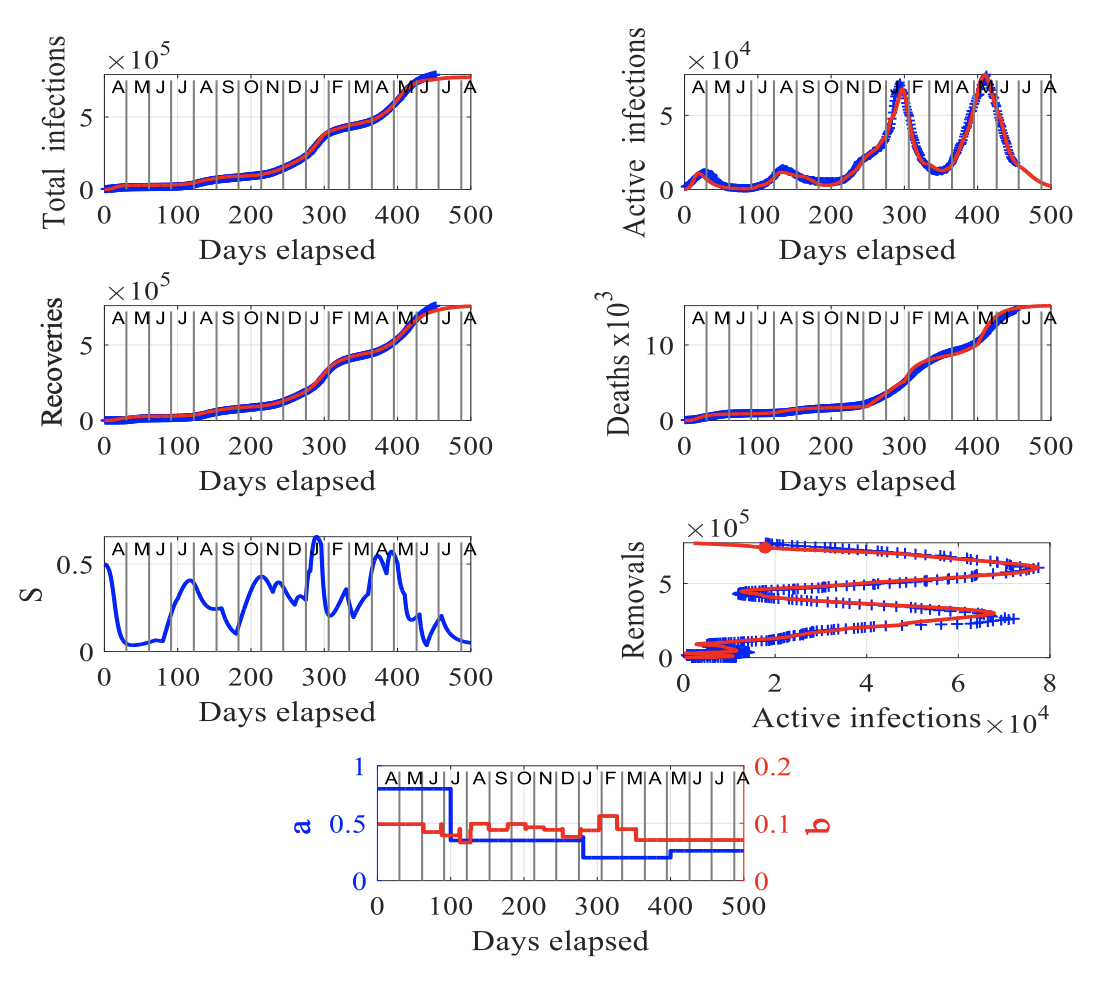}
\caption{Results for Japan: The input parameters to model \eqref{eq_modified_SIR_model} are $f = 5.2 \times 10^4$, $S(0) = 0.5$, $I(0) =3.5673 \times 10^{-3}$ and $R_M(0) =1.0173 \times 10^{-2}$. The letters in the upper axes stand for the names of the months in a year, starting with ``A'' for ``April'' and the red curves are the solutions from our methodology, superimposed to the recorded data shown in blue. To obtain a good fit between the model's predictions and the recorded data, we have assumed rate parameters, $a$ and $b$ that evolve with time as shown in the bottom plot in blue and red, respectively.}\label{Fig_Japan}
\end{figure}

Japan has suffered four distinct waves of infections with small peaks in April and August 2020 and much larger ones in January and May 2021, as shown in the plot of active infections, $I$ in time (days elapsed), in Fig. \ref{Fig_Japan}. The peaks occurred because of the significant increases in the susceptible population (see Fig. \ref{Fig_Japan}) as infected individuals entered and moved about in the country. The third wave-peak in January 2021 resulted in a rapid increase in the number of deaths since the end of 2020,  can be observed in the plot of deaths, $D$ versus time (days elapsed) in the figure. The number of total infections, $I_{tot}$ and recoveries, $R_C$, seems to start stabilizing in June 2021, in accordance with the (non-scaled) susceptibility factor, $S$, which shows a trend to decrease, fluctuating, to very small numbers by August 2021. These results are further corroborated by the plot of deaths, $D$ over time, which stabilize in June and July 2021.

The plot of removals, $R_M$, versus active infections, $I$, in Fig. \ref{Fig_Japan} shows two big horizontal peaks which correspond to  two major outbreaks that occurred  in January and May 2021. Contrary to the temporal behavior of the rate-parameters $a$ and $b$ for the data from Germany (see Fig. \ref{Fig_Germany}), these parameters for Japan in blue and red in the bottom plot respectively, are fluctuating in time with a trend to stabilize both after May 2021.

Similarly to the data for Germany, our methodology for Japan can provide valuable insights into the future trajectory of the spread of the virus in the country, shown in red in Fig. \ref{Fig_Japan}. Assuming no major outbreaks occur between July and August 2021, the situation will improve in Japan. Our forecast suggests that the second outbreak of infections in the country is currently fading out and unless, another outbreak or something in favor of increasing the infection rate does not occur, the situation will improve in Japan. In Subsec. \ref{subsec_forecasting}, we study the case where a major outbreak occurs in early August 2021 in India and  mid-July 2021 in Japan and forecast the number of deaths at the end of August 2021.

%%%%%%%%%%%%%%%%%%%%%%%%%%%%%%

\subsection{India and states}\label{subsec_India}

\begin{figure}[!ht]
\centering
\includegraphics[width=\textwidth,height=11cm]{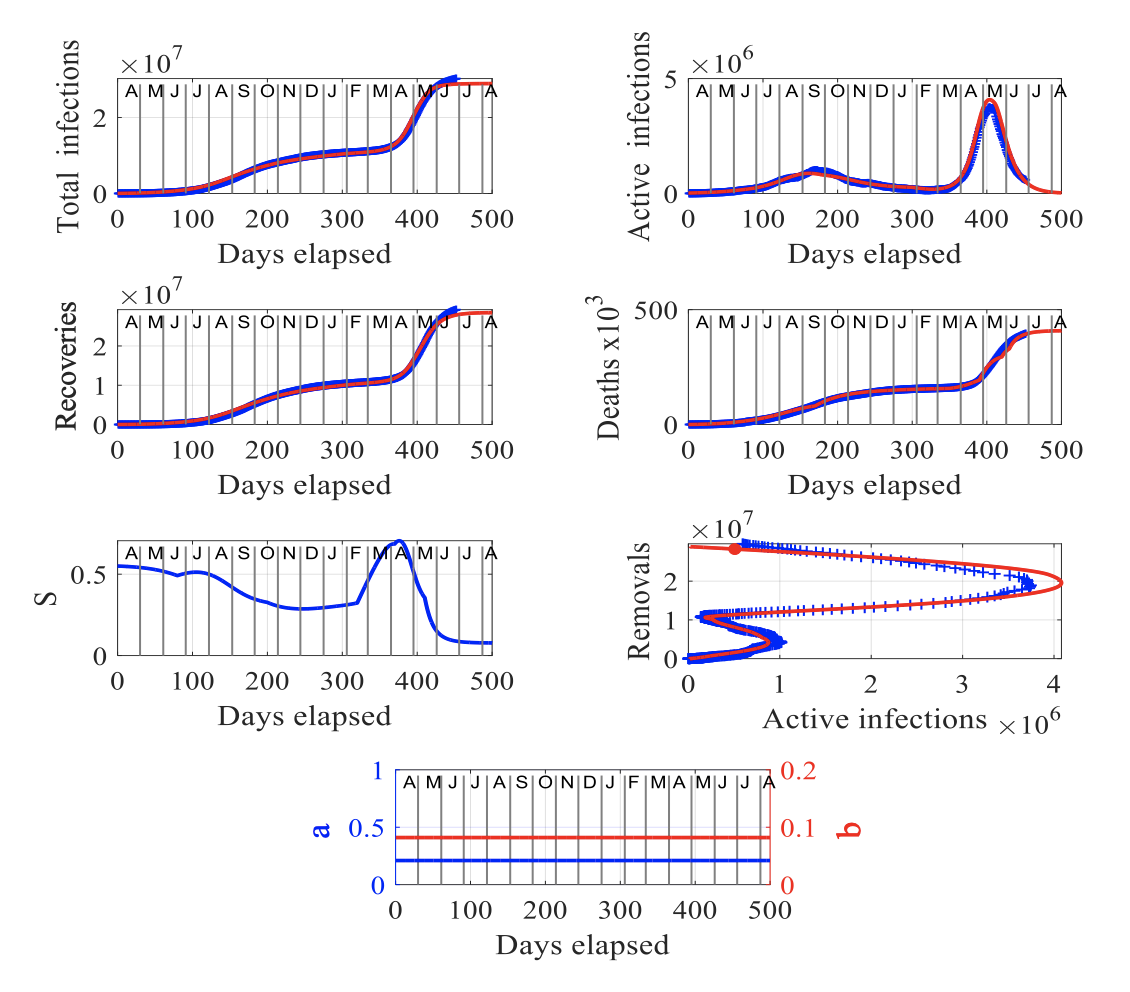}
\caption{Results for India: The input parameters to model \eqref{eq_modified_SIR_model} are $f = 1.2 \times 10^7$, $S(0) = 0.55$, $I(0) = 1.5 \times 10^{-3}$, $R_M(0) =2\times 10^{-5}$. The letters in the upper axes stand for the names of the months in a year, starting with ``A'' for ``April'' and the red curves are the solutions from our methodology, superimposed to the recorded data shown in blue. Here, we have assumed constant rate parameters, $a$ and $b$ to match the model predictions with the published data.}\label{Fig_India}
\end{figure}

Here, we study the data set for India published in \cite{Worldometer_website} and the data sets for Delhi, Maharashtra, West Bengal, Kerala and Karnataka, available in \cite{COVID19-India_website}, in Subsec. \ref{subsubsec_India_states}. India and its states were badly hit by the pandemic and by secondary waves of infections. We start with India, where we had to reset the susceptibility factor $S$ applying a relatively small increment (i.e.,  the order of $10^{-5}$ to $10^{-4}$), from mid-January to mid-February 2021 to match the model-solutions \eqref{eq_modified_SIR_model} to the recorded data set. We increased the infected population between October 2020 and January 2021 at a steady rate of 384 individuals per day to match the recorded data with the model predictions \eqref{eq_modified_SIR_model}. However, from mid-February to mid-May 2021, there was a substantial increase in the number of individuals who became susceptible in the country. This led to the prominent second-wave peak in infections that occurred in early May 2021 as shown in Fig. \ref{Fig_India}. Consequently, a large number of deaths occurred within a one-month period. Compared with the previous 12 months, the number of deaths increased by more than $2\%$ in only one month. During April and May 2021, India was suffering a major second wave of infections with a peak more than 3.6 times higher than the first peak in September 2020. This can be observed in the plot of the total infections over time and more clearly, in the plot of the removals, $R_M$, versus active infections, $I$, in Fig. \ref{Fig_India}, appearing as the two prominent, horizontal, peaks.

Since April 2021, there has been a dramatic increase in deaths in India, surpassing anything that occurred in the country in 2020 due to the virus. In April to May 2021, the virus was not under control and a positive feedback-loop existed in the country in which new infected individuals were moving about, causing other individuals to become susceptible, resulting in more individuals becoming infected and dying as the cycle continued unabated until about mid-May 2021. Towards the end of May 2021, a recovery started appearing in the country, shown in the plot of the susceptibility factor, $S$, versus time (in days elapsed) when the positive feedback loop was broken, since the number of susceptible individuals has continually decreased (see the plot of $S$ versus days elapsed) because of lockdowns and the implementation of vaccination programs.
In the case of India, we have opted for the constant, rate parameters, $a=0.21$ and $b=0.082$ as we have found that for this set of values, the model solutions are very close to the recorded data set. The temporal behavior of $a$ (in blue) and $b$ (in red) can be observed in the bottom plot in Fig. \ref{Fig_India}.

Our methodology applied to the data set for India can provide valuable insights into the future trajectory of the spread of the virus in the country, shown in red in the plots in Fig. \ref{Fig_India}. Assuming no major outbreaks occur between July and August 2021, the situation shows a tendency to improve in the country. Our forecast suggests that the second outbreak of infections in the country is currently fading out and unless, another outbreak or something in favor of increasing the infection rate does not occur, the situation will improve. In Subsec. \ref{subsec_forecasting}, we study the case where a major outbreak occurs in early August 2021 in India and in mid-July 2021 in Japan and forecast the number of deaths in the end of August 2021 in the two countries.

Next, we model the spread of COVID-19 in a number of highly-impacted states in India using our approach in Sec. \ref{sec_methodology}.

%%%%%%%%%%%%%%%%%%%%%%%

%%%%%%%%%%%%%%%%%%%%%%%

\subsubsection{Highly-impacted states in India}\label{subsubsec_India_states}

\begin{figure}[!ht]
 \centering
 \begin{subfigure}[b]{0.48\textwidth}
 \centering
 \includegraphics[width=\textwidth,height=0.22\textheight]{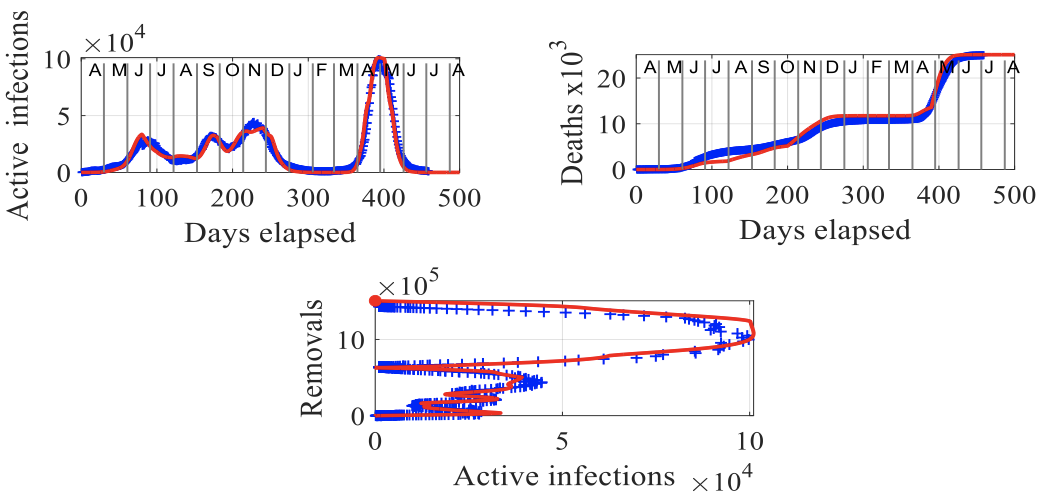}
 \caption{}
 \end{subfigure}
 \hfill
 \begin{subfigure}[b]{0.48\textwidth}
 \centering
 \includegraphics[width=\textwidth,height=0.22\textheight]{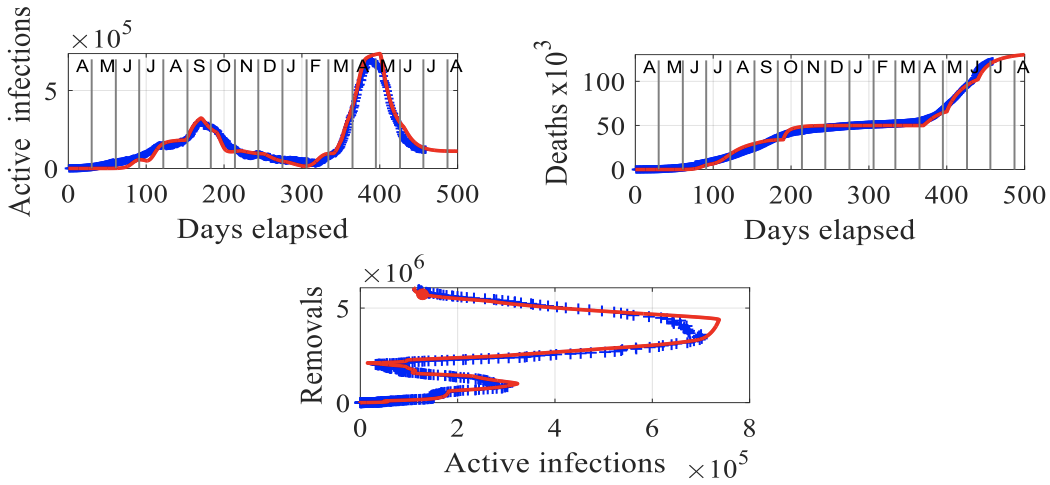}
 \caption{}
 \end{subfigure}
 \hfill
 \begin{subfigure}[b]{0.48\textwidth}
 \centering
 \includegraphics[width=\textwidth,height=0.22\textheight]{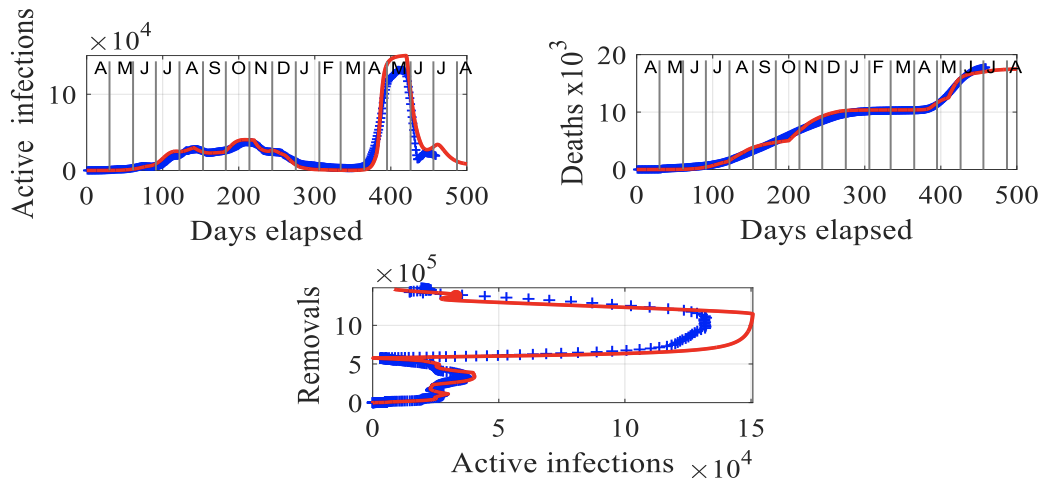}
 \caption{}
 \end{subfigure}
 \hfill
 \begin{subfigure}[b]{0.48\textwidth}
 \centering
 \includegraphics[width=\textwidth,height=0.22\textheight]{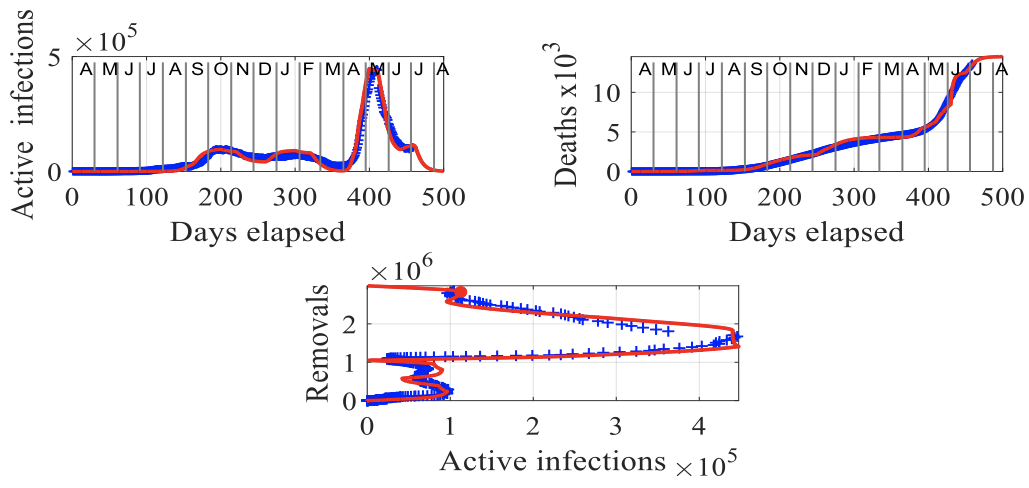}
 \caption{}
 \end{subfigure}
 \hfill
 \begin{subfigure}[b]{0.48\textwidth}
 \centering
 \includegraphics[width=\textwidth,height=0.22\textheight]{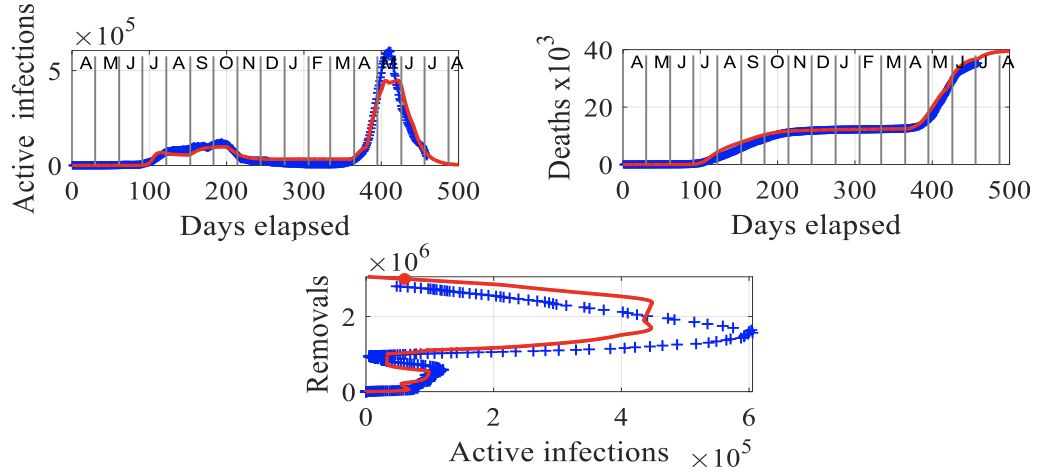}
 \caption{}
 \end{subfigure}
 \caption{Results for Delhi, Maharashtra, West Bengal, Kerala and Karnataka: The input parameters to model \eqref{eq_modified_SIR_model} are: (a) Delhi: $f = 6 \times 10^4$, $S(0) = 0.5$, $I(0) = 2.4 \times 10^{-4}$, $R_M(0) =1.3333 \times 10^{-4}$. (b) Maharashtra: $f = 6.7 \times 10^4$, $S(0) = 0.5$, $I(0) = 4.1791 \times 10^{-4}$, $R_M(0) =8.209 \times 10^{-4}$. (c) West Bengal: $f = 6 \times 10^4$, $S(0) = 0.5$, $I(0) = 5.1667 \times 10^{-5}$, $R_M(0) = 10^{-4}$. (d) Kerala: $f = 3.8 \times 10^4$, $S(0) = 0.5$, $I(0) = 6.2368 \times 10^{-4}$, $R_M(0) = 7.3684 \times 10^{-4}$. (e) Karnataka: $f = 4.1 \times 10^4$, $S(0) = 0.5$, $I(0) = 2.3902 \times 10^{-4}$, $R_M(0) = 2.9268 \times 10^{-4}$. The letters in the upper axes stand for the names of the months in a year, starting with ``A'' for ``April'' and the red curves are the solutions from our methodology, superimposed to the recorded data shown in blue. To obtain a good fit between the model's predictions and the recorded data, we have assumed rate parameters $a$ and $b$ that evolve with time (not shown).}\label{Fig_India_states}
\end{figure}

Here, we focus on the study of the spread of COVID-19 in the badly-hit states of Delhi, Maharashtra, West Bengal, Kerala and Karnataka in India. We are interested in modeling the trajectory of the virus in these states and  contrasting it to that in India. These states have been hit hard by the pandemic and anywhere between one to four outbreaks or surges have been caused during April 2020 and July 2021, depending on the state. We present the results of this analysis in Fig. \ref{Fig_India_states}, where we show the plots of the active infections $I$ and deaths, $D$ over time (in days elapsed) and the plot of the removals, $R_M$ versus the active infections, $I$, starting with Delhi in the upper left corner and ending up to Karnataka in the bottom plot going through the states of Delhi, Maharashtra, West Bengal, Kerala and Karnataka. Focusing first on the plots of infections $I$ and deaths, $D$ in time, we can see that in all five states, there was a series of one to four outbreaks or secondary-waves between June 2020 and February 2021 depending on the state, before the explosion of a big outbreak that occurred around April and May 2021 in India and consequently, in all five states, shown in Figs. \ref{Fig_India} and \ref{Fig_India_states}. These results suggest that the trajectory of the spread of the virus in the five states goes hand-in-hand with that in the country as a whole.

The plots of removals, $R_M$, versus active infections, $I$, in Fig. \ref{Fig_India_states} (horizontal peaks) show the increase in infected individuals that ranges from anywhere between two to six times the number of infected individuals before the outbreak of infections in May and April 2021. Such peaks are markers of major outbreaks in infections in a country and are alarming. This phenomenal increase in the numbers of infected individuals resulted in a dramatic increase in deaths in India and its states, shown in Figs. \ref{Fig_India} and \ref{Fig_India_states} (see plots of $D$ versus days elapsed), surpassing anything that the country has seen in 2020. Consequently, the virus was spreading uncontrollably in India and its states in April to May 2021, due to a positive feedback-loop. During that time, infected individuals were moving about in the country, causing other individuals to become susceptible, resulting in more individuals becoming infected and dying as the cycle continued unabated until about mid-May 2021. The analysis of the results in Figs. \ref{Fig_India} and \ref{Fig_India_states} shows that it was not until the end of May 2021, when a recovery started in the country and in its most impacted states, including those studied herein. That happened as the positive feedback loop was broken, since the number of susceptible individuals continually decreased as a result of lockdowns and ongoing vaccination programs.

%%%%%%%%%%%%%%%%%%%%%%%%

\subsection{Forecasting the spread of COVID-19 in India and Japan}\label{subsec_forecasting}

Our modeling approach in Sec. \ref{sec_methodology} allows to test  the future trajectory of the virus in a community by resetting the susceptibility-factor increment $\Delta S$ (see Eq. \eqref{eq_reset_S}), infection increment $\Delta I$ (see Eq. \eqref{eq_reset_I}) and rate parameters, $a$ and $b$ to emulate the outbreak of secondary waves or surges at specific times. We take this approach to forecast the number of deaths in India and Japan by the end of August 2021, if an outbreak occurs in early August 2021 in India and in mid-July 2021 in Japan.
In particular, both in Japan and India, if there are no more outbreaks or surges in the spread of the virus in July and August 2021, the total number of infections will achieve peak by May 2021, as the susceptibility factors, $S$, approach zero right after (see Figs. \ref{Fig_Japan} and \ref{Fig_India}).

\begin{figure}[!ht]
 \centering
 \begin{subfigure}[b]{0.48\textwidth}
 \centering
 \includegraphics[width=\textwidth,height=0.61\textheight]{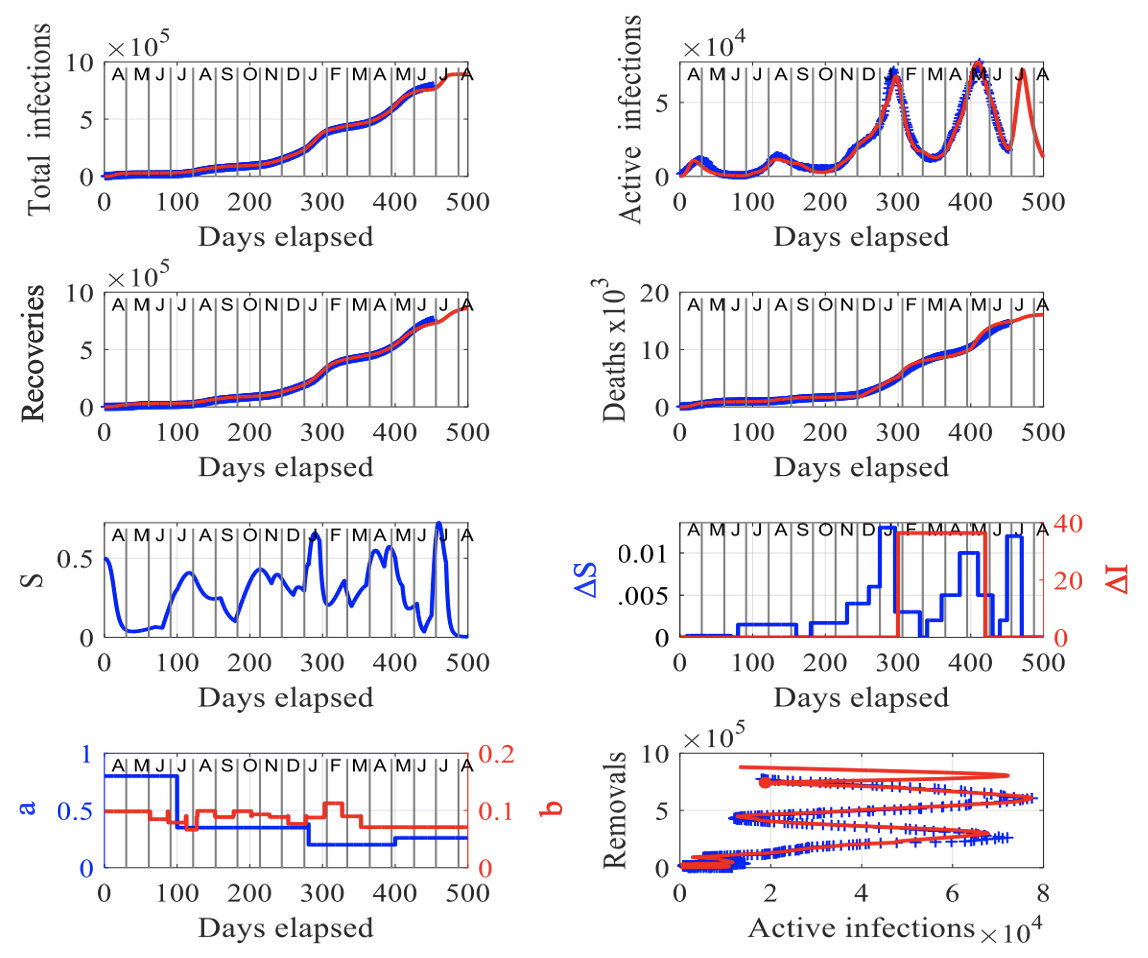}
 \caption{}
 \end{subfigure}
 \hfill
 \begin{subfigure}[b]{0.48\textwidth}
 \centering
 \includegraphics[width=\textwidth,height=0.6\textheight]{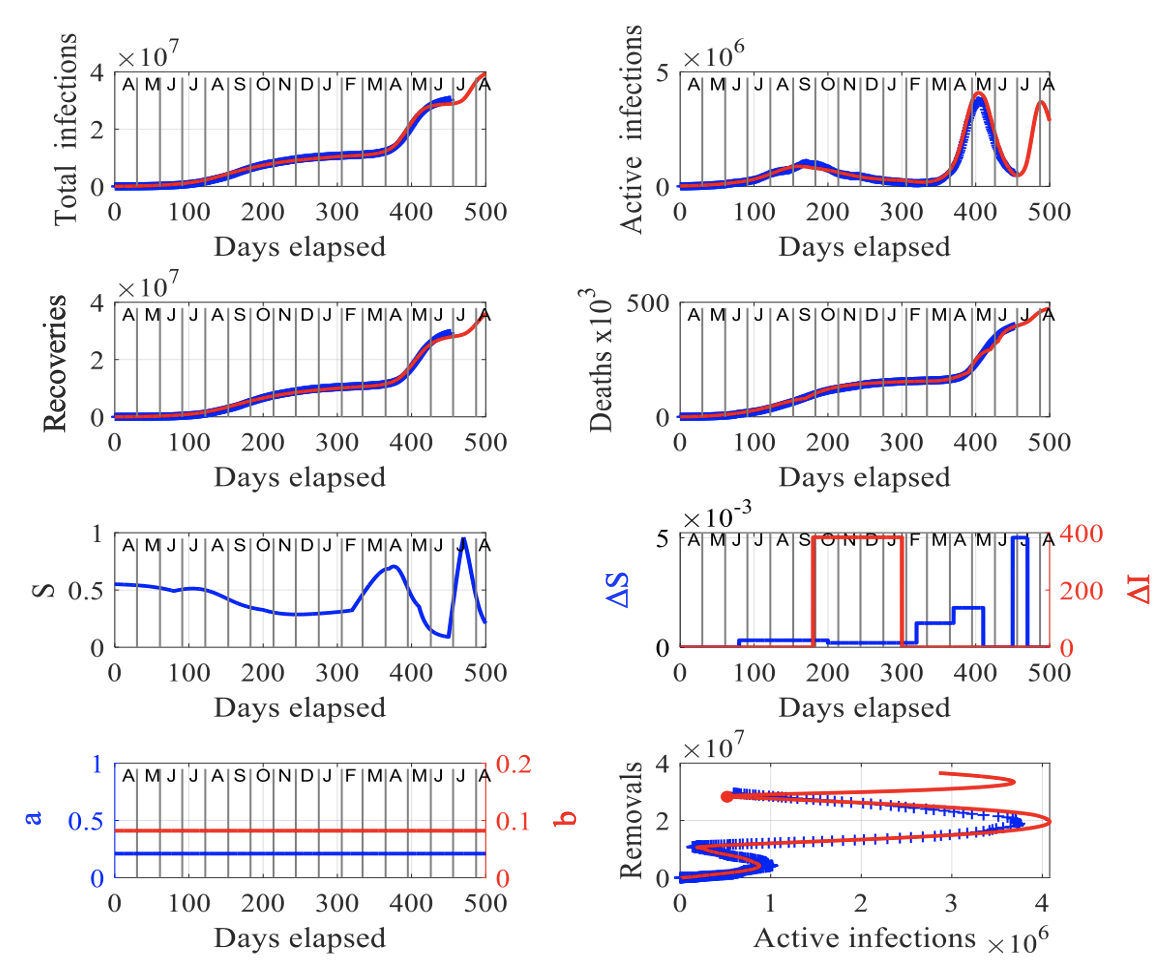}
 \caption{}
 \end{subfigure}
 \caption{The predicted trajectory (in red) of the spread of COVID-19 in India and Japan by the end of August 2021, if an outbreak occurs in early August 2021 in India and in mid-July 2021 in Japan. (a) India: The input parameters are $f = 1.2 \times 10^7$, $S(0) = 0.55$, $I(0) = 1.5 \times 10^{-3}$, $R_M(0) = 2 \times 10^{-5}$. (b) Japan: The input parameters are $f = 5.2 \times 10^4$, $S(0) = 0.5$, $I(0) = 3.6 \times 10^{-2}$, $R_M(0) = 1.02 \times 10^{-2}$. The letters in the upper axes stand for the names of the months in a year, starting with ``A'' for ``April'' and the red curves are the solutions from our methodology, superimposed to the recorded data shown in blue. To obtain a good fit between the model's predictions and the recorded data, we have assumed rate parameters $a$ and $b$ that evolve with time.}\label{Fig_India_Japan_future_trajectory}
\end{figure}

However, for the sake of demonstrating the ability of our proposed methodology to forecast the future trajectory of the virus, we will assume that the vaccination programs and other mitigation measures will not work in India and  the Tokyo 2020 Olympics will finally be held in Tokyo, Japan in late July and early August 2021, as scheduled initially. These are ideal cases to model the consequence of major escalations in infections because of the failure of mitigation measures and  a big sports event by assuming a major outbreak will occur in early August 2021 in India and in mid-July 2021 in Japan, a couple of weeks before the beginning of the games, that is in mid-July 2021. We can model these outbreaks in the data sets by resetting the susceptibility-factor increments $\Delta S$ (see Eq. \eqref{eq_reset_S}), infection increments $\Delta I$ (see Eq. \eqref{eq_reset_I}) and rate parameters, $a$ and $b$ as shown in panels (a) and (b) in Fig. \ref{Fig_India_Japan_future_trajectory} for India and Japan, respectively.

If the vaccination program and other measures do not work in India, then a third wave with an even higher peak in active infections is likely as a result of a new surge in infections in July 2021 (see panel (a) in Fig. \ref{Fig_India_Japan_future_trajectory}). If this surge occurs, it would mean an extra 70 thousand deaths and a staggering extra 10 million new infections in the country. 
If the surge occurred as modeled in Japan because of the staging of the Olympics (see panel (b) in Fig. \ref{Fig_India_Japan_future_trajectory}), then an extra 1000 deaths and a staggering 100000 extra infections would result.

A summary of the number of additional deaths and additional infected individuals by the end of August 2021 is shown in Table \ref{table_forecast_India_Japan} for a surge in infections in India in early-August 2021 because of the failure of mitigation measures, and a surge in infections in Japan at the mid-July 2021, because of the staging of the Olympics.

%%%%%%%%%%%%%%%%%%%%%%%%

%%%%%%%%%%%%%%%%%%%%%%%%

\begin{table}[!ht]
\begin{center}
\begin{tabular}{|c|c|c|c|}\hline
Country & Deaths, $D$ $(\times 10^3)$ & Infections, $I_{tot}$ $(\times 10^6)$ \\ \hline
India & 410 & 30 \\ \hline 
India early-August 2021 surge & 480 & 40 \\ \hline 
Japan & 15 & 0.8 \\ \hline 
Japan Olympic surge & 16 & 0.9 \\ \hline 
\end{tabular}\label{table_deaths_India_Japan}
\end{center}
\caption{Predicted numbers of deaths and active infections in India and Japan by August 2021, if an outbreak in infections occurs in early August 2021 in India (third row) and in mid-July 2021 in Japan (fifth row). The numbers in the second and fourth rows correspond to the cases where no outbreaks would occur in July and August in the two countries and serve as the baselines to compare with the cases of outbreaks in the third and fifth rows, respectively.}\label{table_forecast_India_Japan}
\end{table}

%%%%%%%%%%%%%%%%%%%%%%%%

\section{Discussion}\label{sec_discussion}

COVID-19 impacted deeply on human health, recently due to its frequently changing nature as mutations, such as the delta variant \cite{Kupferschmidt1375} that can spread quickly in different countries. Governments and local authorities have no  choice then, other than taking diverse and adequate countermeasures due to still limited information about COVID-19.
The time dependence of infected individuals in a community can show a multitude of behavior including wavy patterns such as secondary waves and insurgences. To formulate different countermeasures against the spread of the virus, it is desirable to use mathematical models to produce predictive results \cite{liao2020tw,liu2021predicting,scarpino2019predictability,munoz2021sir,tyagi2021mathematical}. These models prove to be effective tools to study, explain and more importantly, forecast the future trajectory of the spread of the virus and of its variants, under different scenarios in states, countries or groups of individuals  \cite{Giordano2020,Hou2020,Anas2020}.

In this paper, we introduce a derived form of an infectious-disease model to calculate the infection curves and  obtain the rate of change of other cases in certain countries and states.  Our approach is based on a modification of the classic SIR model \cite{weiss2013SIR} and is motivated by our earlier research on the spread of COVID-19 in different communities in \cite{cooper2020sir,cooper2020dynamic} and \cite{liao2020tw}. In particular, we introduced a modified SIR model that can account for secondary waves, outbreaks and surges in recorded data sets for communities, either these are countries, states or groups of individuals by  considering recorded data sets published in Worldometers and COVID-19 India websites from April 2020 to July 2021.
Our modeling approach can provide insights into the time evolution of the spread of the virus that the data alone cannot and can be applied to available data sets. As new data are added to the model, one can adjust its parameters and provide best-fit curves between the data and the model-predictions. Hence our modeling approach can provide with estimates of the number of likely future deaths and of time scales for the number of infections. Our model-based analysis and forecast shows that the outbreaks in infections in Germany, Japan, India, Delhi, Maharashtra, West Bengal, Kerala and Karnataka are currently fading out. Unless, further outbreaks or surges in infections do not occur, the situation will improve in these communities, owing probably to the ongoing vaccination programs and mitigating policies implemented by their governments and local authorities.

We also reported on a forecast for the infections and deaths in India and Japan in the end of August 2021, assuming a major outbreak occurs in early August 2021 in India and in mid-July 2021 in Japan. Our model predictions show that if the vaccination program and other measures do not work in India, a third wave with a very high peak in active infections is likely as a result of a new surge in infections in July 2021. If this surge occurs, it would mean an extra 70 thousand deaths and a staggering extra 10 million new infections in the country. Similarly, our model predicts that if an outbreak in infections occurs in Japan in mid-July 2021, an extra 1000 deaths and an extra 100000 infections would result as the country will stage the Olympics in Tokyo between late July and early August 2021.

As far as we know, there are no downloaded data sets for population values needed, hence updates in the proposed model have to be done manually. Data are obtained from the web and entered into a Matlab array. The Matlab script is then executed with parameter updates to give the best fit. The graphical output could then be updated to a webpage. Prediction for future populations can be made for setting values for $S(t)$ and $I(t)$ for $t$ bigger than the time for the last day of data and then these projected values can be adjusted accordingly as new data become available. Hence our model can adapt to the recorded data sets and can be used to explain them and importantly, to forecast the number of infected, recovered, removed and dead individuals, as well as to estimate the effective infection and recovery rates in time, assuming outbreaks occur at specific times. This can be used to forecast the future basic reproduction number and combined with the forecast of the number of infections and deaths, our methodology can assist in the implementation of intervention strategies and mitigation policies to control the infections and deaths in a community. This, supported by the implementation of vaccination programs worldwide, can help to reduce the impact of the spread and improve the wellbeing of people around the globe.

\section{Acknowledgements}
IC is thankful to the School of Physics, The University of Sydney, Sydney,  Australia for the support to complete this work.

%\bibliography{mybibfile_ca5}
\bibliographystyle{spmpsci}      % mathematics and physical sciences

\end{document}